# INCORPORATING AGILE WITH MDA CASE STUDY: ONLINE POLLING SYSTEM

Pritha Guha[1], Kinjal Shah[1], Shiv Shankar Prasad Shukla[1], Shweta Singh[1]

[1]Department of Computer Science and Engineering, Jaypee University of Information Technology, Waknaghat, Distt. Solan (H.P), India.

prithaguha198655@gmail.com, kinjal.93@gmail.com, shivshu@gmail.com, shweta.singh7787@gmail.com

## ABSTRACT

*Nowadays agile software development is used in greater extend but for small organizations only, whereas MDA is suitable for large organizations but yet not standardized. In this paper the pros and cons of Model Driven Architecture (MDA) and Extreme programming have been discussed. As both of them have some limitations and cannot be used in both large scale and small scale organizations a new architecture has been proposed. In this model it is tried to opt the advantages and important values to overcome the limitations of both the software development procedures. In support to the proposed architecture the implementation of it on Online Polling System has been discussed and all the phases of software development have been explained.*

## 1. INTRODUCTION

In MDSD automating the binding between a model and various software systems is just a really small part of the overall problem. It is accepted by the industry and researchers that MDA [1] systems solve a small portion of systems development and will typically suffer from the "stale design" and performance problems. The path forward for software development is perhaps to use MDA[2] as a prototyping exercise, but the real productivity gains will come from ever increasing productivity tools (like better GUIs, APIs and programming models) and increasing metadata. MDSD is helpful specifically in the case when the industry has some predefined tools like templates, GUIs are ready. If the industry has to make the software from the scratch it has to go through the diagrams of model and paper works will lead to greater delay when industry has to meet the specific deadline and the client is very eager to know the progress of his functioning project from the software industry because ultimately client is only concerned about the working code of the project. The client is not interested much in the diagrams and paper documentation work. Whereas in Extreme programming methodology [3] of agile [4][5] software development, it is focused more on coding, design, testing and deadline of the project. Methodologists often describe Extreme Programming as the stereotypical agile method [16]. When any software is being developed there are many basic risks which are associated with software development namely periodic slips in which the industry denies the client to deliver the working software when deadline can’t be met by the industry persons. These periodic slips born the other risk namely cancellation or rejection of the project. The client is always interested in giving project to the company which can meet the deadline of working software with flexible in dynamic changes in requirements [15]. If these requirements are being fulfilled by the industry in the context of the development of the software the client is ready to pay much more to the best of the price of the software to the developing firm [15]. The other risk is the software production goes into infinite testing because the software is tested on each different demand of the client so it mostly frustrates the developers as well as testers. This will lead to another major problem called defect rate which become very high due to changing requirements from the client side. Every time the programmers will have to write new and new functions for adding new features with different testing associated with it. The frequent changing requirements by clients leads the misconduct in software production with the quitting scenario of good programmers and developers because after one saturation level the programmers hate to work on project and hence intended to quit the work. During the development process sometimes the developers adds some funny features to make the client happy but those features don't help much to the client to make the money and it is referred as false feature in the software. The associated risks are very crucial to solve at the time of working software launching process [13]. Agile methods deal well with unstable and volatile requirements by using a number of techniques of which most notable are: low ceremony documents, short iterations, early testing, and customer collaboration [16]. The form of agile software development method is also a process of thought originality [18]. The extreme programming address the problems discussed here very efficiently and in fruitful manner. It acts on its five values which are communication, feedback, respect, courage and simplicity, which accelerates the production work rather than involving much in the documentation work to satisfy the client in the fruitful manner. But this is only successful for developing a small scale project. Thus both the aspects of the software development are the best in their areas according to project definition, deadline, and budget given by the client to the software industry.

## 2. RELATED WORK

In this area the related work up till now has been done towards Change oriented adaptive software engineering by using agile methodology which is referred as Cognizant Feature Driven Development (CFDD). CFDD does not cover the entire software process but it will highlight on the design and building phases. The model mainly handles the incoming change in the requirements very efficiently. In CFDD the change comes from the change source from there it is passed to the change handling phase where change is being examined and necessary decision is taken in the aspect of the development. After taking the decision the change is passed to the quick development phase from there it will be developed in terms of real code and then passed to the Final Review and release stage. This stage will launch the change from the paper to the real working code. When in the future if the same kind of the change occurs or comes, it is handled much faster using the CFDD model [14].

In the other area of this field the customer satisfaction are mainly focused at the time of software development. This model has classified the customer satisfaction variables in to three parts namely Customer Changes, Requirement Changes, Agile Practices. The Requirement changes phase has introduced four crucial phases which are shortly after development started, during development (main phase of the product development), shortly before the end of development, after development was completed. The Agile practices are subdivided in to six phases which are iterative development, Work Climate, Final Product Adaptability, Continuous Customer Integration, Efficient Execution and Willingness to adapt the change [15].

The other area in the software engineering goes towards the Model Driven Architecture which introduced ARCHMDE approach. This approach is union of architecture centric and model-driven paradigms can facilitate the automation of transformation process and capitalize the reuse in MDE approach. The main objective of ARCHMDE is independence of the software architecture from any platform implementation. This architecture split the Platform

Independence Model (PIM) in to Architecture Independent Model (AIM) and Architecture Specific Model (ASM). These both models have not any technology specific implementation information. Platform Specific Model (PSM) is derived from the Architecture Specific Model (ASM) from which code can be generated [19].

Other research area on the MDA has classified the MDA technology in to four phases namely Project Initiation Phase, Software Development Process Analysis and Selection Phase, MDA Support Phase, SDP Execution Phase. The MDA Support Phase is further divided in to following subsections namely Platform identification and specification, Modelling Language Identification and specification, Transformation Identification and specification and Tool Selection Phase. The SDP Execution Phase is further classified to the SDP Execution Initiation, Component Identification and overall planning; Iterative Development Cycles includes Cycle Planning, Component Development, Modelling, Verification, Model Transformation, Coding/Testing, Integration and deployment, Process and Quality Review followed by Final Release respectively [20].

Papers such as [25] [26] [27] have explained the challenges related to Model Driven Development which are classified as Understanding and managing the interrelations among partly redundant artifacts, Comparing and merging different versions of models, Difficulties with transformations of models (to code or for other types of models). Rampant round-trip problems, Model-level debugging is not supported by tools, Combination of graphical and forms-based syntaxes with text views is not well supported, Moving complexity rather than reducing it, More expertise required.

In the proposed system architecture the challenges to the Model Driven Architecture is solved up to certain level with the help of merging the MDA technology with the Agile Software Methodology's Extreme Programming method. In the given architecture the whole software development process will be following the specific phases which accelerate the launching of the software fasters as well as encourages the clients for changing the requirements if any very frankly even in development phase and teaches the employs of the industry for the best reusability of the existing code for adding the newer features as per the demand with in the specific deadline.

## 3. MODEL DRIVEN ARCHITECTURE

It forms the backbone for any successful software intensive system and it is considered as first-class elements in the system design and modeling. Architecture is the primary carrier of a software system's quality attributes such as performance or reliability [19]. Model-driven development (MDD), also often called Model Based Software Engineering (MBSE), or Model driven Engineering (MDE) is a software development method that focuses on creating models, or abstractions, as first class development artifacts of software and transformations of those models to produce the source code [21] [23]. Model-Driven Software Development is a new software development paradigm for distributed project teams involving 20+ people, with roots in software product line engineering, which is the discipline of designing and building families of applications for a specific purpose or market segment also it is a software development approach that aims at developing software from domain-specific models. The promise of "MDA" is that a model will be used to model the application data, and then the logical model is bound/converted/mapped to the various underlying languages and data structures [19]. By using the model as the source of metadata for the application, the model won't ever get out of date either.

## 3.1 Advantages of MDA

Productivity increases because the programmers only have to model the system and do a few customizations. If you decide to change from platform that shouldn't be a problem but it is only the theoretical aspect. If you only have some pieces of your application modeled you'll be able to do prototyping. All you have to do is to push the 'Generate code' button.

Also, hand coding of model behavior is bypassed and the total focus is on modeling i.e. make the Metamodel instead of focusing on design.

These Meta models are also known as platform specific model (PSM) which provide abstraction. They always reflect the current baseline by separating the business rules from the diverse implementation environment.

MDA tries to separate the platform characteristics [24]. It [6] [7] also provides the documentation feature. In the MDA approach the documentation at a high level of abstraction will naturally be available. But the MDA tools which are nowadays on the market aren't able to generate 100% code. So you'll always have to code after the generation process (like complex business rules). MDA separates the two aspects about the development of the project. One is Platform Specific Model (PSM) and other is Platform Independent Model (PIM) [19] [23] [24]. After finalization of the PSM or PIM the model is finally given to the developers for coding so the notion can come to the real world based on the architecture [22] [23].

## 3.2 Disadvantages of MDA

If you want to use a relatively new technology, like JSF (Java Server Faces), then you must keep in mind that the MDA tool is the developer for you. So it takes some time before the producer has implemented a new technology into their MDA tool.

One of the major drawbacks is that MDA does not provide a standard for the specification of mappings different implementation of mappings can generate very different code or models. MDA systems almost invariably suffer from the "design documents collect dust on bookshelves" problem, despite best attempts of the tools and organization to stop the natural entropy.

The problem is that MDA systems only solve about 5% of the problem. 95% of the efforts of a software development project understand customer requirements, creating the architectures and designs to solve the requirements, understanding / creating the semantics of the components in new and legacy systems. Automating the binding between a model and various software systems is just a really small part of the overall problem.

It is agreed that MDA systems solve a small portion of systems development and will typically suffer from the "stale design" and performance problems. The path leads to the software development is perhaps to use MDA as a prototyping exercise, but the real productivity gains will come from ever increasing productivity tools (like better GUIs, APIs and programming models) and increasing metadata.

MDA does not prescribe any specific development process for enacting model transformations in the context of a software development effort. MDA offers no guidance as to the process (phases, activities and roles) to be used. Since MDA does not prescribe a specific development methodology, each MDA-based development project has to define its own process, or choose a process from the extremely sparse set of MDA based methodologies available [20] [21] [22].

## 4. EXTREME PROGRAMMING

Working Software is the primary concern in context to the client [14]. The industry adopting the extreme programming has project manager, analyst, designer, coder, and tester. The extreme programming adopting industry must have strong programmers and testers which provides the stability in terms of software development. The team members in the extreme programming must be good programmers who can perform one or more responsibility so that the development work gets done faster. In Extreme programming the industry having expertise developers has the best velocity for the software development. This will support the industry in succeeding in the market as well as getting good projects faster and faster like Infosys, Patni Computers, Wipro, Satyam Computers, etc.

### 4.1. Advantage of Extreme Programming:-

Testing:-During this phase the testers and developers find the defects having separate test for the product immediately so the code may get ready for the future use [10] [14].

Simple Design:-It means the code must have less classes and methods as possible with having no duplicate logic with mentioning every important state which is useful for programmers after the code has been passed from the all kind of testing [10] [14].

On site customer:-This phase is very crucial for client as well as industry point of view because this phase acts like bridge between clients and programmers because the client guides the programmers according to his test for the successful completion of project leading no disputes at both the ends [10] [14].

Pair Programming:-This phase accelerates the success of project because it is always to do programming for every module of the project in group or pair rather than individual [10].

Refactoring: - It means the maintenance of the code should be  in a good, understandable as well as in no duplicate form [10].

Small Release:-It is the practice to release the midterm product in context of full product demanded by the client because after testing the midterm product the client can make sure the industry is going according to his test in developing the product as well as he can guide much better to the industry regarding his test in future remaining product [10] [14].Ultimately client always eager to know the progress regarding his project given to the industry [15].

Continuous Integration:-When the tester or developer find any defect or bug in the product using continuous integration they can optimize the size of code so that bug detection and removal may be done very efficiently [10].

Metaphor:-It represents the logical as well as analytical view between business people and client to convince and reveal "what it is being tried to do and in which direction the  flow would be" [11].

Software Quality:-This phase always practices to make the software quality high as in terms of functionality according to client's test. This helps the programmers to handle any kind of projects with the valuable experiences in doing programming in the various projects to improve software quality. Ultimately this phase improves the quality of software industry in launching various products because the programmers are getting trained at each and every stage [11][14].

Extreme Usability:- In Extreme programming this phase helps the developer to protection against dissipation of energy because of most of the code are ready and they can utilize that code in the development of the product ,hence they need not to code extremely every time once the code or template is available like "Login Module". They can reuse the code. This saves a lot of time of developer to think in the aspect of some innovative ideas as well as logics in making and optimizing the code of product in much efficient manner [12][14].

Periodic slips:- In Extreme programming helps the industry for shorter release of functional software. At the time of releasing the customers are contacted and ask for the highest priority features in the software and based on the priority the next release of the working software are launched [13].

Cancellation of Projects:- Extreme programming saves the project against the cancellation with the help of periodic releases [13].

Business Misunderstood Problem:- Concerning customers will lead the industry to protect against the business misunderstood problem which will help the customers to interact actively with the industry [13].

The Defect Rate:- The developers also get interest in doing the project from the client so defect rate generally goes down day by day and quality of the industry on launching the product will improve [13].

False Features Problem:- Extreme programming always focuses on the highest priority tasks so false features are not prioritized during the development of the software. It gives the freedom to the developers and testers to give their feedbacks upon the release time and cost of the software which will helpful for interaction with the clients via the business people [13].

### 4.2. Disadvantages of Extreme Programming

The problem with the Extreme programming is that it is best suited for single project, developed and maintained by a single team. It cannot be implemented in the system where developers don't work well with each other and like to work by their own.

XP will not work in an environment where a customer or manager insists on a complete specification or design before they begin programming.

Several potential drawbacks which include problems with unstable requirements, so no documented compromises of user conflicts, and lack of an overall design specification or document.

Due to Extreme Programming Software industries need to save a large amount of unprecedented code which sometimes effects the certain amount of profitability of the industry because using this code the industry persons develop new features from the old developed one but sometimes it may possible that previously designed code may never get used and just become the head ache to save it carefully which leads to large and large data centers having cost effective infrastructures as well as hardware includes a noticeable amount of power supply consumption on daily basis.

## 5. PROPOSED ARCHITECTURE

Both the software development processes described above has limitations and many drawbacks. This shows the need of new model for developing software which is in standardized form and can be implemented in all type of environments. Here, a new architecture has been proposed which implements the Extreme programming in model driven software development lifecycle. Different phase of the architecture been proposed as follows.

## 5.1 Phase 1:- Requirement and Domain Analysis

This is the very first phase of this architecture. In this requirements from the user are taken into consideration and according to those requirements domain analysis is done. The requirements may be functional or non-functional. After the proper negotiation with the customer a requirement specific document is developed. Documentation is very important aspect when are developing software for large organization. Proper authorization and reliability is provided by this. Henceforth the whole development process is divided into multiple domains. These domains are then passed to next stage of the architecture as shown in the figure 1.

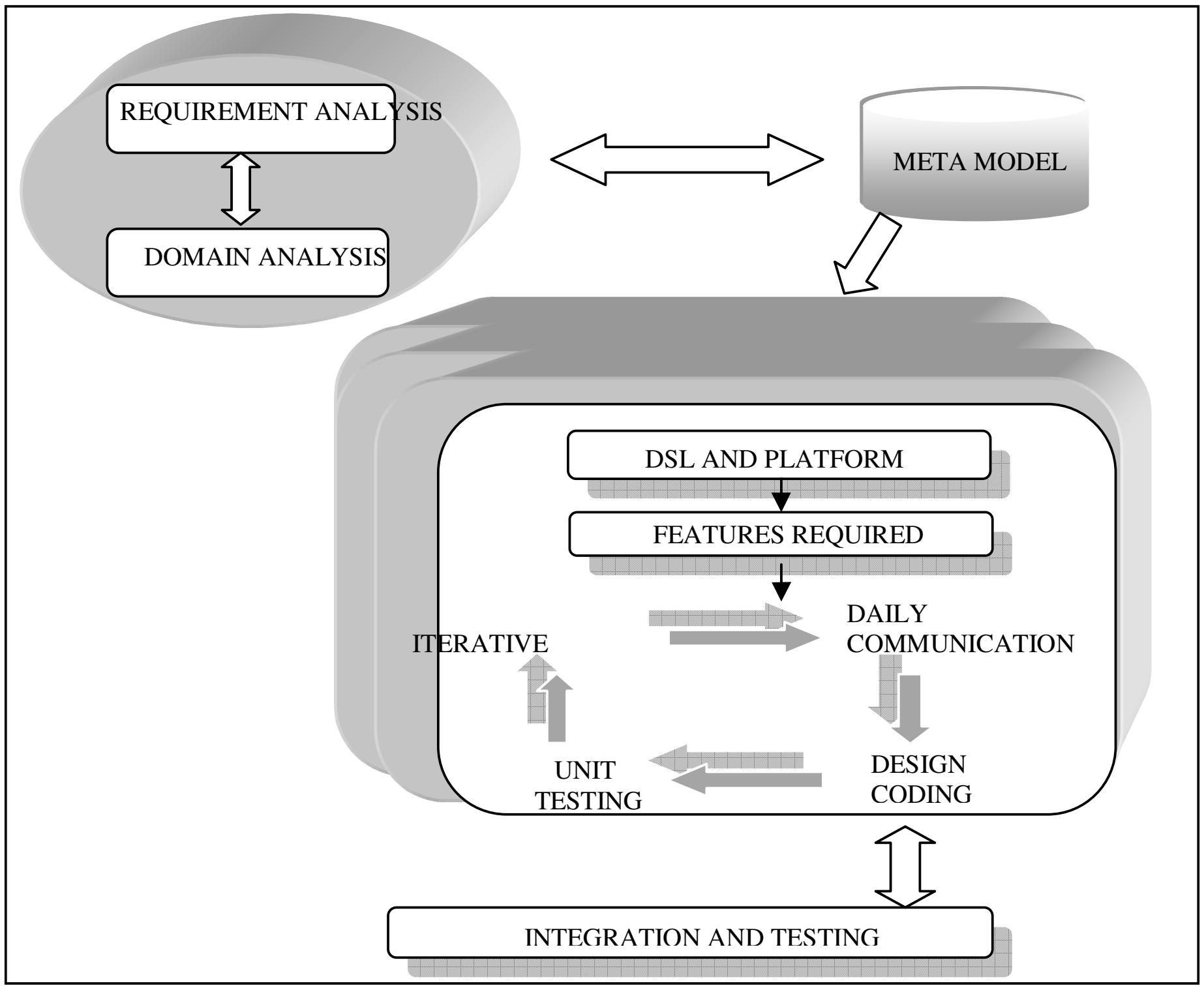

Figure 1.Hybrid Architecture

## 5.2. Phase 2:- Metamodel

During this stage a Metamodel is developed according to the requirement analysis and domain analysis. The model is then confirmed with customer for the final design consideration. Till this phase customer can add its requirement and get a complete overview of the development process going to be used. The Metamodel design is also a part of the documentation and negotiation is done here also with the customer. Henceforth customer can interfere and view the functionality till this stage. Once the Metamodel is being developed main part of the development process starts which is explained in next section. The divided domains now work individually in Extreme programming manner.

### 5.3. Phase 3(a):- Manager

One important task of the manager is that it act as central interface between all the domains and help them to communicate with each other. Firstly the DSL and the platform is decided for that particular domain then after that it is been communicated with the manager so that each domain`s platform must be compatible to each other.

### 5.4. Phase 3(b):- Working Domain

All requirements and features are identified and implemented. An iterative Planning procedure is opted for development of that domain part with daily communication, coding and testing. Once planning is done it is communicated to all team members and they design and code accordingly. Unit testing is done on that code and if error occurred planning is done again until a bug free application is achieved. Domain applications are now passed to next stage.

### 5.5. Phase 4:- Integration and Testing

This is the last phase of this architecture in which integration of all the domain applications is done and testing is performed again. This is the last phase as if all domain applications are running efficiently with each other our software has been developed then. As in previous stage domains are checked for compatibility there is less chance for failure in integration and testing capitalized. This all phases as shown in Figure 1 follow the algorithm as explained in the Table: 1 as shown below.

Table: 1 Algorithm for Hybrid Architecture

| | |
|---|---|
| Step 1 | int RD=0,M=0,D=0; |
| Step 2 | Do Requirement and Domain Analysis; |
| Step 3 | RD=1; |
| Step 4 | Meet the client and ask for his/her/its requirements with the budget of the Software; |
| Step 5 | Do the negotiation with the client regarding the requirements and budget in the aspects of the platform/domain as well as cost for developing the Software; |
| Step 6 | if (client agrees with the negotiation in all aspect of domain and cost area) |
| Step 7 | Do prepare the customer requirement specific model for the Software; |
| Step 8 | Done; |
| Step 9 | Else |
| Step 10 | break; |
| Step 11 | end if |
| Step 12 | Done; |
| Step 13 | Done; |
| Step 14 | RD=2; |
| Step 15 | if (RD!=2) |
| Step 16 | go to step:2; |
| Step | Else |
| 17 | |
| Step 18 | Do Enter in the Meta Model Phase and Prepare Meta Model according to Requirement and Domain Analysis; |
| Step 19 | M=1; |
| Step 20 | Show the Meta Model to the Client for make sure the final design of the software's flow and ask him regarding any of the changes he wants; |
| Step 21 | if (Client wants change/suggest any change in the flow of the software) |
| Step 22 | Note down the changes/advices; |
| Step 23 | Else |
| Step 24 | Freeze the final requirements for any change regarding the software and cost from the client side as well as industry side on the legal document; |
| Step 25 | end if |
| Step 26 | end if |
| Step 27 | done; |
| Step 28 | M=2; |
| Step 29 | if(M!=2) |
| Step 30 | go to step:15; |
| Step 31 | Else |
| Step 32 | Do enter in the Development phase which includes Manager who decides the DSL and Platform used for developing the software and Required features according to client's test and Working Domain includes Iterative planning, Daily Communication, Design Coding, Unit Testing; |
| Step 33 | D=1; |
| Step 34 | Done; |
| Step 35 | end if |
| Step 36 | if(D!=1) |
| Step 37 | go to step:29; |
| Step 38 | Else |
| Step 39 | Do Module Integration and Perform Unit Testing; |
| Step 40 | if(find some bugs or error) |
| Step 41 | go to step:32; |
| Step 42 | Else |
| Step 43 | Launch the product after communication with the higher authority of the industry; |
| Step 44 | end if |
| Step 45 | done; |
| Step 46 | end if |

### 5.6. Advantages of Proposed Hybrid Model

It can be used for both small and big infrastructure. XP was implemented in small scale project whereas this architecture can be implemented in both small and large scale project with multiple domains.

Here whole process proceeds proper documentation and complete negotiation with the customer. The customer can add its requirements in the middle of the software development lifecycle.

In MDA each domain is bound to use particular DSL and platform whereas in this each domain has to decide its own DSL and platform according to its domain requirement after communicating with the manager despite of using a common one. Due to use of multiple domains which works parallel, the whole procedure can be completed in least amount of time.

## 6. CASE STUDY

### 6.1 Phase-1

In phase 1 as discussed earlier it needs to be listed the user requirements and according to that the domain analysis should be done. In this scenario our client is the government which wants from the industry to develop the ONLINE POLLING SYSTEM and users are voters of every state. The primary goal is to fulfill the requirements regarding that the domain and requirements of the polling software should be listed. User requirements as given as follow- the software must show the Population of the state, total voters, candidate information, status if voter is alive or dead. Authentication/validation of voter is done by their unique Voter ID which will be displayed along with their personal information. Requirements also contain a fully organized database which containing audit logs for each domain also it will be checked that voter cannot vote twice. Regarding the domain analysis it is concluded that there are 26 different domains having different population of voters, priorities and issues.

Domain analysis- Client stated that out of 26 domains 10 are very reach and have no cost limitation for higher security and these domains have higher populations so they need better technology and security. Similarly the demand is of  6 domains which have less financial support but very sensitive and require higher security at a lower cost and remaining 10 domains are of intermediate level in terms of both finance as well as security. This is the very first phase of this architecture.

In this requirements Client also wants that these requirements may be different for different states but finally it has to be made the central server which is having the knowledge of each and every voter of all 26 domains. Finally all local domains must be connected with the central domain along with the above requirements because these requirements are common in each domain. With proper communication with the client the requirement and domain analysis is done. Further it should be progressed with developing a Meta model about our development phases.

### 6.2 Phase-2

In this phase, a platform independent analysis model is defined through analyzing the requirements phase. System functionalities are described in this Meta model while maintaining traceability to the requirements. Here the main idea is that the functionality of the software is designed by the central government only. So the client should be in contact and do negotiation with central government only. If domains of the software need any changes then it will report the manager and there is no interference of the client. Developers may use appropriate model elements stored in a model repository to produce this. This model is not the final model but forms the foundation for producing the final version. It is finalized only after approval of client. The general plan may be reviewed if necessary. The Meta model is finally developed using tools which work on above constrains. Constraints, preconditions, post conditions, and invariants are defined using UML and OCL mechanisms. The product of this activity is the main framework of the system.

### 6.3 Phase-3

In Phase 3 based on approval on Meta model by client satisfactorily the work will be divided according to each domain. In each domain the Data Specification Language and Platform is decided based on Budget and security issue. These requirements guide the developers whether it should be worked on Linux, Windows kind of Platform and DSL like Php, java, Perl, Python,c#.net etc. After finalizing the DSL and platform, the features for using that kind of language and platform are required to be installed like if PHP is decided then XAMPP server is required, if J2EE is advised then apache tomcat and net beans on the system are required, also Glassfish application server, JBoss server for J2EE can be used based on the specification. Other web application servers like IBM's Web sphere which is too much expensive that must be decided upon the budget of the client.

Even BEA's Web logic application server can be used so similarly if .Net technology for developing the web-application is being used then Microsoft Visual Studio 2005 or 2008 or 2010 with MSSQL server 2005 database is required. If firm decides to use Perl language for developing web-application then Perl Interpreter must be there on the system as well as .Net supports Perl language so GUI using different tools like ADOBE Dream Viewer, Net beans, and Microsoft Visual Studio can be developed easily. If Perl, Python and Java languages are being used for development of web-application then Linux platform can be more suitable because Linux has inbuilt support for Perl and Python Interpreters and as everyone knows Java is platform independent Language. Here industry mostly focuses on the open source technology like LAMP (Linux Apache MySQL PHP), Django which is open source Application framework written in Python and follows Model View Controller Architecture loosely. Panther tool which is an open source software for building n-tier database applications. Similarly More-Motion Application Studio which is Java Based Open source for developing Web-application on Linux Platform rather than using expensive IBM's Web-sphere, Microsoft's Visual Studio on Windows Platform which are costly for lower budget client due to license and Software costs.

After deciding the features the Manager's Role arises which has control over all the domains. Now manager allots different module to the different groups of persons of same domain like if the industry has a team of 10 members then manager will allot to design of pages to 2 persons, design database to 3 persons, linking and database queries to 2 persons, create view to 3 persons including mother tongue and manager will be responsible for synchronization between different domains via daily communication report basis. After daily communication the Design and coding will be started based on DSL and Language Specification and features. After the coding the unit testing is performed in the very urgent manner to cope up with the deadline. Now during unit testing and coding if the team members stuck at any problem then manager solves that problem and also it suggests the iterative solution to that problem like rather than showing only age of VOTER it must be applied its DOB also so client can verify the age. Also he suggests that the database team should utilize MySQL database, PostgreSQL database rather than SQL server 2005 due to the fastest seeking time as well as they are open source. This all aspects come under iterative planning. Manager also keeps track of the work progress of one domain must not be lagging behind from the progress in another domain for a day or two. Thus Manager maintains Synchronization between all the phases.

### 6.4 Phase-4

In this phase the output of each domain is taken as input and adds all those modules together one by one into single working software. Here different domains have their own budget and security limitations so considering this it is needful to integrate all the domains combined to one single central server via some high standard technology like .net which supports 44 different development languages. Here with similar Integration tool different domain codes in to one platform can be integrated and perform parallel testing on each and every module. During testing if bugs are found, error and compatibility issue it should be tried first to resolve it by own otherwise it should be returned back to the phase 3 communicate with the manager and according to problem it must be resolved .Thus integration of whole unit and testing can be done by applying all possible inputs as well as it must also be checked that in heavy load/traffic condition this software works fine.

## 7. CONCLUSION

In this paper a new architecture for a software development process using the Extreme programming methodology of agile software development in core Model Driven software development. Here the limitations of both the models have been tried to remove and the proposed model can be used for large organizations as well as for small organizations. With the help of the given architecture the Manager can interact with the developers, testers as well as the people associated with the task of integration on daily basis in a very productive and well-organized manner by just providing them Data Specification Language with the features demanded by the clients. The Manager in the architecture also keep track of the necessary deadlines on the release of the product so the Business people associated with the Requirement and domain analysis may not get fail in their commitment. The manager in the proposed system architecture plays very much crucial role to interact with the business people and the developers, testers and integrators for the cost of the software as well as the commitment given to the client. This will help to prevent any kind of confliction that can occur between the production people and the business people. This will be supportive to the industry to create the goodwill of it among the clients as well as the other competitor industries regarding the commitment and the launching of the product on time. This model is very much flexible in changing the requirements by the client to the best of its extent.

In this paper the case study of online polling system software has been taken which can be developed using proposed architecture. If MDA is selected during software development for a large organization, the client needs to give his requirement in the Meta-Model phase. If he wants to change his requirements in the development phase, it will not permissible for the client to do so. On the other hand if agile model is being used for smaller organization, client can change his requirements during development phase also but using our proposed architecture the client is free to change his requirements in the Meta Model phase as well as in the Development phase. The proposed architecture trains well the developers especially for reusability of the existing code for tackling the changing requirements of the clients on same module with some different features required. This architecture creates a good bonding between the limitation of agile software development as well as Model Driven Architecture during the development of the online polling system software.

## Authors

**Pritha Guha** received her M.Tech degree in Computer Science and Engineering from Jaypee University of Information Technology, Waknaghat, Distt.Solan in 2011.She graduated from Gyan Vihar School of Engineering and Technology, affiliated with Rajasthan University in 2008. Currently she has been selected by Wipro Technology India Ltd. Her area of interests are Data Mining, Data Structures, Software Engineering, Operating Systems.

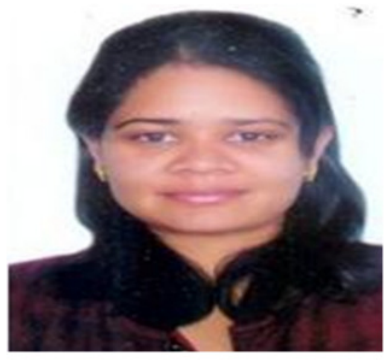

**Kinjal Shah** received his B.E degree in Computer Engineering from A. D. Patel Institute of Technology, New Vallabh Vidyanagar, Distt. Anand (Gujarat) in 2009. He is currently pursuing MTECH degree in Computer Science and Engineering at Jaypee University of Information Technology, Waknaghat, Distt. Solan-173215. His research interest includes Cloud Computing, Cryptography and Network Security, Multimedia Data Transmission in ad-hoc networks, Software Engineering.

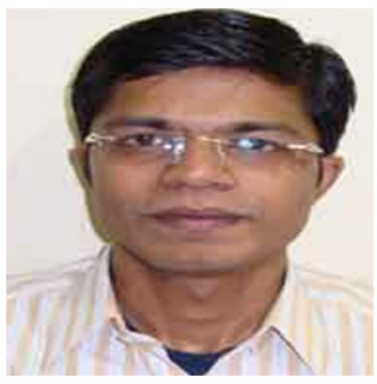

**Shiv Shankar Prasad Shukla** received the Diploma's degree in Computer Application from Government Polytechnic Mining Institute, Dhanbad (Jharkhand) in 2007. He got Bachelor's Degree in Computer Technology from Priyadarshini College of Engineering; Nagpur (Maharashtra) in 2010.He is currently persuing the M.Tech degree in Computer Science and Engineering at Jaypee University of Information Technology, Waknaghat, Distt. Solan-173215. His research interest includes Fraud Detection, Cloud Computing, Cryptography and Network Security, Flaming in Social Networking and Software Engineering**.**

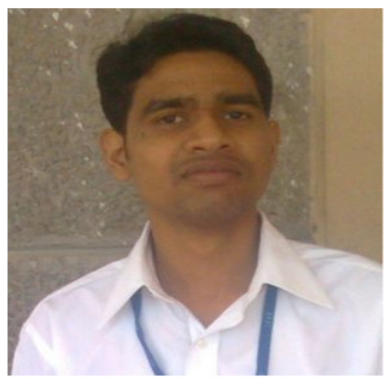

**Shweta Singh** received her B.TECH. degree in Computer Science Engineering from Subharti Institute of Technology and Engineering in 2010.Currently she is pursuing M.TECH degree in Computer Science and Engineering at Jaypee University of Information Technology, Waknaghat, Distt. Solan-173215. Her areas of interest are Theory of computation, Advanced Algorithms,Software Engineering, Computer Networks.

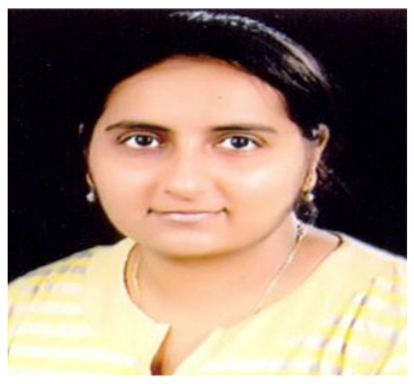